%% file: sample-sigconf.tex
\documentclass[sigconf]{acmart}
\usepackage[ruled]{algorithm2e}
\usepackage{algorithmic}
\usepackage{booktabs}
\usepackage{makecell}
\usepackage[group-separator={,}]{siunitx}
\usepackage[export]{adjustbox} % loads also graphicx
\usepackage[title]{appendix}
\usepackage{lipsum}% http://ctan.org/pkg/lipsum
\usepackage{graphicx}% http://ctan.org/pkg/graphicx
\usepackage{multicol}% http://ctan.org/pkg/multicols
\usepackage{geometry}% http://ctan.org/pkg/geometry
\usepackage{blindtext}

\setcopyright{rightsretained}
\begin{document}
\title{Attack Surface Metrics and Privilege-based Reduction Strategies for Cyber-Physical Systems}
%\titlenote{Produces the permission block, and copyright information}
%\subtitle{Extended Abstract}
%\subtitlenote{The full version of the author's guide is available as
 % \texttt{acmart.pdf} document}

\author{Ali Tamimi}
\orcid{0000-0001-7099-8337}
\affiliation{
  \institution{Washington State University}
  \streetaddress{School of Electrical Engineering \& Computer Science}
  \city{Pullman}
  \state{Washington}
  \postcode{99164}
}
\email{ali.tamimi@wsu.edu}

\author{Ozgur Oksuz}
\affiliation{%
  \institution{Washington State University}
  \streetaddress{School of Electrical Engineering \& Computer Science}
  \city{Pullman}
  \state{Washington}
  \postcode{99164}
}
\email{ozgur.oksuz@wsu.edu}

\author{Jinyoung Lee}
\affiliation{%
  \institution{Washington State University}
  \streetaddress{School of Electrical Engineering \& Computer Science}
  \city{Pullman}
  \state{Washington}
  \postcode{99164}
}
\email{jinyoung.lee@wsu.edu}

\author{Adam Hahn}
\affiliation{%
  \institution{Washington State University}
  \streetaddress{School of Electrical Engineering \& Computer Science}
  \city{Pullman}
  \state{Washington}
  \postcode{99164}
}
\email{a.hahn@wsu.edu}

% The default list of authors is too long for headers.
%\renewcommand{\shortauthors}{A. Tamimi et al.}

\begin{abstract}
Cybersecurity risks are often managed by reducing the system's attack surface, which includes minimizing the number of interconnections, privileges, and impacts of an attack.  While attack surface reduction techniques have been frequently deployed in more traditional information technology (IT) domains, metrics tailored to cyber-physical systems (CPS) have not yet been identified. This paper introduces attack surface analysis metrics and algorithms to evaluate the attack surface of a CPS. The proposed approach includes both physical system impact metrics, along with a variety of cyber system properties from the software (network connections, methods) and operating system (privileges, exploit mitigations).  The proposed algorithm is defined to incorporate with the Architecture Analysis \& Design Language (AADL), which is commonly used to many CPS industries to model their control system architecture, and tools have been developed to automate this analysis on an AADL model. Furthermore, the proposed approach is evaluated on a distribution power grid case study, which includes a 7 feeder distribution system, AADL model of the SCADA control centers, and analysis of the OpenDNP3 protocol library used in many real-world SCADA systems. 
\end{abstract}

%
% The code below should be generated by the tool at
% http://dl.acm.org/ccs.cfm
% Please copy and paste the code instead of the example below.
%



\maketitle

\input{samplebody-conf}

\bibliographystyle{ACM-Reference-Format}
\bibliography{bibfile1}

\end{document}

%% file: samplebody-conf.tex
\section{Introduction}
Modern cyber-physical systems (CPS) increasingly depend on large, complex software platforms to monitor and control complex environments. Often the physical domains being controlled are large, distributed systems (e.g., air traffic control, electric power grid), that depend on a significant number of sensor measurements and actuators, requiring critical centralized platforms where operators can monitor and control these environments. However, while the cybersecurity of these systems is increasingly important, their complexity presents many challenges to the risk assessment process.

Both the bulk transmission and distribution segments of the electric power grid provide examples of this.  The bulk transmission system is controlled from a large number of control centers, each with a Supervisory Control and Data Acquisition (SCADA) server that communicates with a large array of distributed sensors (e.g., current/voltage transformers) and actuators (circuit breakers) deployed within geographically disperse substations. The SCADA server collects data from the substations, which consist of IEDs, RTUs, sensors, relays, circuit breakers, while the energy management system (EMS) performs higher-level analysis and optimization algorithms (e.g., state estimate, AGC) to inform operator decisions. Furthermore, distribution-level control centers also utilize SCADA servers and distribution management systems (DMS) to perform a unique set of control applications (e.g., Volt-VAR control, load flow).  

Recent events have demonstrated that control centers are key targets of attacks, as the 2015 Ukrainian attack targeted distribution-level control centers and in 2018 reports surfaced that attackers targeted U.S. power grid control centers~\cite{ukraine:eisac} \cite{ta18074a}. Furthermore, these events also demonstrated that attacks to the control center could result in a significantly greater impact than those to individual substations.  While significant work has focused on assessing system vulnerabilities from malicious data to applications (e.g., state estimation, AGC), there has been insufficient work exploring the vulnerability of such systems to attacks that manipulate the control platform (e.g., software exploits).  This concern has been recently validated through the discovery of software vulnerabilities in key protocols used to support the wide-area communication between these devices~\cite{DNP3vulns}. Such vulnerabilities could allow an attacker to escalate privileges from a poorly protected substation or pole-top device into the control center, which would provide the attacker with the ability to manipulate large amounts of control and sensor data.

Furthermore, while attacks to control systems platforms are an increasingly serious threat, these systems are also seeing a continually expanding attack surface. On the transmission side, substations are adding additional PMU devices and new wide-area communication networks~\cite{4735961}. On the distribution side, substations increasingly contain remotely controlled devices (e.g., switches, transformer taps), while systems are increasingly interconnected with more vulnerable pole-top devices (e.g., voltage regulators), consumer-owned distributed energy resources (smart inverters), and smart meters~\cite{pserc:ami}. 

To protect against false data injection attacks, researchers have demonstrated that control applications can be enhanced by robust control techniques~\cite{7438916}, along with traditional efforts such as cryptographic authentication (e.g., HMAC, digital signatures). However, to protect against software vulnerabilities, protection methods commonly include (i) protecting memory (ALSR, DEP), and (ii) deploying mechanisms to improve the isolation between software components (e.g., privileges, visualization, trusted execution environments).  While these mechanisms are increasingly important for the protection of complex cyber-physical systems, there is currently limited work exploring how the cyber-physical properties of the system influence the allocation of these security mechanisms in a manner to best protect the system.

To address this challenges, this paper introduces the following contributions. First, it presents attack surface metrics and analysis algorithms for cyber-physical systems which incorporates the impact of an attack, the degree of system connectivity, and a variety of cybersecurity properties of the system's software. Second, it demonstrates the proposed metric based on a cyber-physical AADL system model which enables broader adoption to many different systems and introduces a tool to automates this analysis~\cite{github:aadltool}. Third, it presents a use-case based on a cyber-physical distribution power system model using a seven feeder model control using the OpenDNP3 protocol library ~\cite{opendnpthree}. Fourth, it demonstrates the proposed techniques on various system privilege models to demonstrate techniques to strategically reduce system's attack surface.  

\section{Related Work}
The challenge of growing system attack surfaces is well defined within security literature~\cite{GeerJr}. Early attempts to define metrics for attack surface included work by \cite{rasq} which introduces a Relative Attack Surface Quotient (RASQ) metric that enables multiple system configurations to be measured against each other, such as sockets, named pipes, RPC endpoints, running services, weak ACLs, and user accounts. In \cite{5482589}, the authors extend this formal model for attack surface based on system privileges, access rights, and methods which were then evaluated on a various open-source software package.

Similarly, many research efforts have introduced attack surface evaluation techniques to help identify more secure system configurations and assist in the comparison of various system architectures. For example, work in \cite{Kurmus_attacksurface} utilized graph-based models to analyze the attack surface of Linux kernel configuration while incorporating both the lines of code and call graphs for various functions.  In \cite{6980440} the authors present a system-level view to compare the attack surface of two different certificate validation approaches, DANE and X.509.  Attack surface metrics have also been explored to provide techniques to reduce privileges in Android apps by minimizing the permissions given to each app. In \cite{Szefer:2011:EHA:2046707.2046754} the authors explored the attack surface of modern hypervisors to evaluate the security of cloud-based applications. From a CPS perspective, work in \cite{6025254} explored graph-based models and algorithms to explore the attack surface of various key information objects used to control the grid, while in \cite{7013366}  the authors propose a technique to reduce the attack surface by dynamically controlling network paths.  

%Relationship between Attack Surface and Vulnerability Density: A Case Study on Apache HTTP Server

In addition to attack surface efforts, many researches have explored both vulnerabilities and mitigations of power center control center applications through false data injection attacks. For example work in~\cite{Liu:2011} and many others have explored false data injection in state estimation algorithms, while in~\cite{6740883} the authors explore techniques to prevent false data injection in AGC. Further work has explored the attack impact to economic dispatch algorithms ~\cite{8023151}, power system market operations~\cite{6074981}, and smart meter deployments~\cite{ForemanG16}. In \cite{cpac2016}, the authors present a cyber-physical access control solution to mitigate threats in cyber-physical environment. They provide information flow analysis and logic-based policy control to stop harmful operation in the industrial control systems. While there is significant work in these areas, there has been little work explore the security of the software platforms that execute these algorithms and must connect with various untrusted systems and networks.  

This paper will utilize AADL to provide a standard model for the cyber security and extends other research efforts which use AADL to define and analyze system security properties. A report by \cite{cmu_aadl} discusses a variety of use-case applications to improve cyber-physical security through the implementation of AADL, including modeling techniques for threats and access levels. The authors in \cite{5306062} have demonstrated that AADL can be used to perform modeling-checking on information flows to protect both data confidentiality and integrity, work in \cite{milsaadl} has demonstrated multiple independent levels of security (MILS) validation using AADL. Furthermore, \cite{aaspe} introduced attack tree and impact analysis tools for AADL  along with a standard for a Security Annex. 

\subsection{Comparison to Related Work} 
Our work expands upon the software-based attack surface metrics introduced by \cite{5482589} which aggregates the risk from the various methods, channels, and data within software to compute a quantitative score. Specifically, it explores $\emph{entry-exit points of the system}$ through the $\emph{methods}$ (e,g., API) that allow the attacker to exchange data directly with the system.  The $\emph{channels}$ (e.g., TCP) that are used to send and receive data between the attacker and the system, and  $\emph{data}$ (e.g., file) that is used to exchange information between the attacker and the system. A unique metric is identified for each factor based on its $attackability$, which is defined as damage potential-effort ratio based on each resource.  The $\emph{damage potential}$ can be seen as the level of harm the attacker can cause to the system in using the resource in the attack. $\emph{Effort}$ measures the attacker's difficulty to acquire the necessary access rights to be able to use the resource in an attack. The higher the ratio, the higher the resource's contribution. Damage potential is defined as a method's privilege (root, authorized user, unauthorized user), a channel's type (TCP, UNIX SOCKET, SSL) and a data type (file, registry) while the effort is defined as the access rights of a method, channel, and data.

\begin{table}
\footnotesize
\caption{Expansion of work in  \cite{5482589}. to CPS}{
\centering
    \begin{tabular}{|p{1cm}|p{3.1cm}|p{3.2cm}|} \hline
      \textbf{Resources}        & \textbf{\cite{5482589}} &  \textbf{CPS Unique Properties}    \\ \hline
 Data  & Data damage potential based only on datatypes (e.g., files, db)           
       & Data has quantifiable system impact to physical processes \\\hline
               
Channel & Unknown number of external connections and unclear impact of data manipulations
        & Typically well defined interconnections and impact metrics for data in messages \\\hline
              
Method & Focuses on methods within single software platform, not system-level impacts
        & Analysis incorporates propagative impact of attack to other processes within a system \\\hline
\end{tabular}}
\label{tbl:diff}
\end{table}

\subsection{Contributions CPS Attack Surface} 
This work expands upon previous work to define CPS attack surface metrics, as explained in Table~\ref{tbl:diff}. First, in a cyber-physical system, the impact of manipulations to data (e.g., sensor measurements, actuator commands) should be quantified based on their manifestation within a physical domain. Therefore, we utilize physical system impact metrics to evaluate the damage potential of attacks. Second, compared to pure software-based attack surface metrics which cannot make assumptions about the number of interconnections or the operating environment, a CPS will typically have a defined set of devices or external resources that it interacts with. Third, a CPS commonly must aggregate many software platforms within a single environment to perform various functions including (i) monitoring and state estimation, (ii) control and feedback, (iii) human operator interfaces, (iv) network communication, (v) data storage and archiving. While only some of these functions perform direct cyber-physical functions, all are necessary to support the system's operation and therefore, contribute directly or indirectly to the attack surface.

\section{Model for CPS Attack Surface}
\label{sec:metrics} \label{sec:def}
In this section, we introduce a system model and then define attack surface metrics that incorporate both the cyber and physical system properties. The metric will emphasize the criticality and vulnerabilities of the interconnectivity between a system and its external connections. The metrics will also explore the negative impact that an attack could have on the physical process. 

For the cyber-layer model, it will assume systems are defined using AADL \cite{AADL}, which is a highly standardized system modeling language that is widely used within many CPS industries, including both aerospace and automotive \cite{OSATE}. In AADL, the system model incorporates both hardware and software components. Hardware components include memory, processors, buses, and devices, while software components include processes, threads, and data.  A component is a set of software (processes, threads, thread groups, data, subprograms) and execution platform (processor, memory, devices, bus) mechanisms \cite{AADLBook}. A component can also be a subsystem which can include a composite of these sub-components. A {\it system} then defines the set of connections between the components and also introduces a mapping of various software components to the physical devices  (e.g., hardware components). Furthermore, {\it connections} can be used to define communications between systems, subsystems, and devices, while {\it flows} can represent the data that traverses from a source to a sink. Moreover, we could define {\it security properties } in the AADL model and assign them to different components.

\subsection{Definitions} 

\begin{definition} [System] A system, $s$, is defined similarly as in AADL, where it contains a set of software components, hardware components, and subsystems. A system can be divided into a set of trusted $s_t$ and untrusted $s_u$ components.
\end{definition}

\begin{definition} [Attack Paths] The attack paths for trusted subsystem $s_t$ is defined as $ap(s_t) = \{c_1, ..., c_n \}$ and represents the set of connections that connect $s_t$ to the components of the untrusted subsystem ($s_u$). 
\end{definition}

\begin{definition}[Privilege] \label{def:lr} Most operating systems enforce access control by defining $privileges$ and assigning them with access rights to the various objects (programs, data) on the system. Therefore, a system will be defined with a set of privileges $V(s) = \{v^{s_i}_1, \ldots, v^{s_i}_n\}$, where each privilege, $v^{s_i}_j = d_1,\ldots,d_k$, which defines what data items ($d_1,\ldots,d_k$) a specific process can access. Furthermore, we define a mapping $Priv : P  \rightarrow V $ that assigns a privilege to each process. 
\end{definition}

\subsection{Attack Surface Metrics} 
The attack surface metrics will be computed for the trusted subsystem, $s_t$, based on the number of available attack paths ($ap(s_{t})$) that connect the trusted system to its untrusted components. The total attack surface measurement is defined as $TASM(s, ap(s_t))$, which is calculated as the summation of the individual attack surfaces of each connection within $ap(s_t)$ as defined below.  

\begin{equation}
TASM(s,  ap(s_t)) = \sum_{i \in ap(s_t)} AS_C(i) + AS_P(i)
\end{equation}

The metric for each connection incorporates the attack surface contributions of the connections ($AS_C(i)$) and the process ($AS_P(i)$) in $s_t$ that is connected to that system. Each $AS$ metric is defined a value $imp \times exp$, where $imp$ (impact) is defined based on attackers ability to impact the physical system based on the manipulation of some data and the $exp$ (exposure) is defined based on how vulnerable the component is to attack. 

\subsubsection{Data} As defined in the previous sections, the impact of an attack to a component is evaluated based on the set of data it processes, stores, or transmits. However, the impact of an attack would be quantified differently across different cyber-physical system domains, furthermore, it can also be quantified different within the same domain. For example, the following list provides a brief survey of papers analyzing various cybersecurity attacks to different applications within the smart grid where the attack impact is measured through different variables, including costs, estimated states, or operational values (e.g., load, frequency, voltage).\\
\indent \textit{ 1)} Power system loss of load (MW) \cite{mani:vulnassess}\\
\indent \textit{2)} Generation cost (\$/MWh) \cite{8023151}\\
\indent \textit{3)} DC State Est. (volt., power) \cite{Liu:2011}\\
\indent \textit{4)} Locational marginal price (\$/MWh) \cite{6074981}  \\
\indent \textit{5)} AC State Est. (volt. mag., phase) \cite{6672638}   \\
\indent \textit{6)} Stability Factors (freq., volt., angle) \cite{6787098}\\

% \begin{tabular}{p{3 cm} p{.1 cm} p{4 cm}} 
% \textit{1)} Power system loss of load (MW) \cite{mani:vulnassess}&& 
% \textit{4)} Locational marginal price (\$/MWh) \cite{6074981}  \\
% \textit{2)} Generation cost (\$/MWh) \cite{8023151} &&
% \textit{5)} AC State Est. (volt. mag., phase) \cite{6672638}   \\
% \textit{3)} DC State Est. (volt., power) \cite{Liu:2011} &&
% \textit{6)} Stability Factors (freq., volt., angle) \cite{6787098}   \\
% $\quad$ & $\quad$
% \end{tabular}\\

To address this challenge, we introduce a generic attack impact metric,  $imp(d)$, which must be defined based on domain-specific methods to quantify the impact of the manipulation of data $d$. A case study in Section 4 will be provided to demonstrate how the loss-of-load metric can be used to analyze the attack surface of a distribution grid's control center.

\subsubsection{Connection}\label{conn_as}  For each connection, the attack surface will analyze the impact of the data that traverses that connection along with the exposure of that channel to various network-based threats. The connection's attack surface is then defined as $AS_C(i) = exp_c \times imp_c$, where $exp_c$  and $imp_c$ represent the exposure and impact of that channel. 

A connection's exposure is be determined by whether it is physically exposed (e.g., wireless) and also based on the extent to which it includes security mechanisms to protect the communication (e.g., IPSec, TLS). There are four factors considered for evaluation of connection's exposure including encryption algorithm, key exchange mechanism, hash function and type of connection. Table \ref{tab:conexp} shows different values of the factors.

\begin{table}
\footnotesize
\caption{connection's exposure factors and scores}{
\centering
 \begin{tabular}{|l c|l c|l c|l c|} \hline
 \multicolumn{2}{|c|}{\textbf{Encryption }}&\multicolumn{2}{|c|}{\textbf{Key}}&\multicolumn{2}{|c|}{\textbf{Hash}}&\multicolumn{2}{|c|}{\textbf{Media}} \\
  \multicolumn{2}{|c|}{\textbf{Algorithm }}&\multicolumn{2}{|c|}{\textbf{Exchange}}&\multicolumn{2}{|c|}{\textbf{Function}}&\multicolumn{2}{|c|}{\textbf{Type}} \\
   \multicolumn{2}{|c|}{\textbf{}}&\multicolumn{2}{|c|}{\textbf{Mechanism}}&\multicolumn{2}{|c|}{\textbf{}}&\multicolumn{2}{|c|}{\textbf{}} \\ \hline
 \multicolumn{1}{|l}{\textbf{Alg.}}&\multicolumn{1}{c|}{\textbf{Sco.}}&\multicolumn{1}{|l}{\textbf{Mech.}}&\multicolumn{1}{c|}{\textbf{Sco.}}&\multicolumn{1}{|l}{\textbf{Func.}}&\multicolumn{1}{c|}{\textbf{Sco.}}&\multicolumn{1}{|l}{\textbf{Type}}&\multicolumn{1}{c|}{\textbf{Sco.}} \\ \hline
 %Alg. & score&Mech. & score&Func. & score&Type & score\\ \hline
 AES & 3&2048 bits & 3&SHA-2 & 3&Wired & 1\\
 3DES & 2&1024 bits & 2 &SHA-1 & 2&Wireless & 0 \\
 DES & 1&512 bits & 1&MD5 & 1& & \\
 Null&0&Null&0&Null&0& & \\ \hline
\end{tabular}}
\label{tab:conexp}
\end{table}
 
Using the scores, we define $t_{c}$ in Equation \ref{eq:tc}. Values from Table \ref{tab:conexp}  are assigned to parameters based on the communication protecting mechanisms. The $encryption$ parameter defines the algorithm used for encrypting the data. The values of the parameter for AES, Triple DES (3DES), and DES are 3, 2, and 1 respectively. If there is no encryption algorithm, the value of the parameter is set to zero. The $key\_exchange$ parameter shows that what kind of prime is used for exchanging the key. We give the values 3, 2, and 1 to 2048-bit, 1024-bit and 512-bit prime respectively. If there is no key exchange mechanism, the value is zero. The $hash\_function$ shows what kind of cryptographic hash is used for generating the signature. The value 3, 2, and 1 are assigned to SHA-2, SHA-1, and MD5 hash algorithm respectively. If there is not any hash algorithm, the value of the parameter is zero. Finally, if we use a wired media for communication, we assign 1 to $media\_type$ parameter; Otherwise, the value is set to zero. If the mechanisms are more secure, the value of $t_{c}$ is larger. It results in a smaller value of $exp_c$ which is calculated in Equation \ref{eq:expc}.
\begin{equation}
t_c = encryption+key\_exchange+hash\_function+media\_type
\label{eq:tc}
\end{equation}
\begin{equation}
exp_c = \frac{1}{1+t_c}
\label{eq:expc}
\end{equation}

In addition to the exposure metrics, methods are also needed to evaluate the impact of an attack that can tamper any data that traverses this channel and would be vulnerable to a false-data injection (FDI) attack. Therefore, the impact of this attack is defined as the set of data that traverses the channel.

\begin{equation}
imp_c = \sum_{d \in conn_i(d)}  imp(d) 
\label{eq:impc}
\end{equation}

\subsubsection{Process} The attack surface for a process will incorporate a variety of factors including its privilege, the set of exploit mitigations enabled to protect it, and the criticality of the data accessible by it ($v^{s}_{p}$). The attack surface for a $\emph{process}$ is defined similarly to the definition for a $\emph{connection}$, $AS_P(i)  = exp_p \times imp_p$, where $p$ is the trusted process in the connection $i$.

The exposure of a process is based on multiple factors, including the set of protection mechanisms enabled by the operating system along with the complexity of the process. The protection mechanisms will include common best-practices implemented by operating systems to protect processes from exploitation. These include:
\begin{enumerate}
\item \textit{Address Space Layout Randomization (ASLR)} protects memory against buffer overflow attacks by randomizing the memory location of processes. By using this mechanism, the attacker is unable to find correct address space location necessary to control of a process execution  \cite{Bojinov2011}.
\item \textit{Data Execution Protection (DEP)} marks certain pages of memory non-executables so that if the code that is potentially injected into memory, such as through a buffer overflow, cannot be executed \cite{DEP}. 
\item \textit{Control Flow Integrity (CFI)} protects against invalid execution traces for an application by restricting the flow-control of the application to the known paths of a Control-Flow Graph (CFG) \cite{Abadi}. 
\item \textit{Code signing} utilizes a certificate-based digital signature to sign executables and scripts to guarantee executed code is protected from any change or corruption since it is signed \cite{cooper2018security}. \item \textit{64-bit process} provides improved security as it increases the address space for a process (compared to 32-bit architecture) and makes other security techniques, such as ASLR, more effective.

\end{enumerate}
 Using these factors, we evaluate $t_p$ in Equation \ref{eq:tp}.
\begin{equation}
t_p = ASLR+DEP+code\_signing+ 64\_bit+CFI
\label{eq:tp}
\end{equation}

For each mechanism, if it is implemented the value in the formula is 1. Otherwise, it is equal to 0. Therefore, $t_p$ is a number between 0 and 5. Then, $exp_{c}$ is calculated in Equation \ref{eq:expp}.
\begin{equation}
exp_p = \frac{M}{1+t_p}
\label{eq:expp}
\end{equation}

Where $M$ is the number of methods of the process. We could analyze the code of a process using the tools such as cflow \cite{cflow}, and evaluate the number of methods.
%Additional operating system-specific mechanisms can  also be incorporated into this. The total mitigation score is defined as $||m||$, where $m = \{0,1\}^n$ representing whether each of the $n$ feasible exploit mitigations are enabled. 
%\begin{equation}
%exp_p = ||m|| \times \color{red} METHODS\color{black} 
%\end{equation}

While an attack to a channel (Section\ref{conn_as}) only allows the manipulation of any data communicated over that channel, an attack to a process could allow manipulations of all data defined by the privilege($v^{s}_{p}$) of that process. Therefore, a process attack's impact should include all data accessible by that process privileges and is therefore defined as follows. 

\begin{equation}
imp_p = \sum_{d \in v^{s}_{p}}  imp(d) 
\end{equation}

\subsection{AADL-based Attack Surface Analysis Algorithm} 
In this section introduced a propose an attack surface analysis algorithm based on the previously defined metrics. The algorithm assumes the system is modeled utilizing AADL and requires that the user specifies the system model, attack paths, and the security properties of processes and communications. It also assumes the model includes a number of other defined system variables, such as the physical system impacts (that depends on the open switches on the system) and the number of methods for subprograms. While these are not natively defined in AADL but could be added through the AADL properties file.

The Algorithm is implemented as a plug-in which developed by Eclipse modeling framework. The plug-in uses the AADL model as input and categorizes the components of the model including substations, processes, etc. based on their attack paths. For each component, the security properties are extracted. If the type of component is "connection", the security properties are including impact, encryption algorithm, key exchange mechanism, hash function, and media type. By using Equation \ref{eq:tc} and \ref{eq:expc} the attack metric for the component is calculated. If the type of component is "process", the security properties are ASLR, DEP, code signing, 64 bit, and CFI. The attack metric is calculated based on security properties of the process and Equation \ref{eq:tp} and \ref{eq:expp}. Finally, by summing up the attack surface metric of the components of a path, we evaluate the path attack surface metric. 
Algorithm \ref{alg:mainalg} shows how the total attack surface metric is calculated.

\begin{algorithm}
\caption{Find Total Attack Surface Metric (TASM)}\label{alg:mainalg}
\centering
\begin{algorithmic}[1]
\STATE calculate\_metric(AttackPath[] $ap$)\{
\STATE $TASM \leftarrow 0$
\STATE \For {each $path$ in ap}
{
    \STATE $ path\_metric \leftarrow 0$
    \STATE \For {each $component$ in path}
    {
    	\If {$Type\_of(component)=="connection"$} 
        {
        	\STATE $properties \leftarrow Extract(component.properties)$
           \STATE $tc \leftarrow  component(encryption) + component(key\_exchange) + component(hash\_function) + component(media\_type))$
			\STATE $ expc \leftarrow 1/(1+tc)$
            \STATE $ component(metric) \leftarrow expc * component (impact) $
        }
        \If {$Type\_of(component)=="process"$} 
        {
        	\STATE $properties \leftarrow Extract(component.properties)$
            \STATE $tp \leftarrow  component(ASLR) + component(DEP) + component(code\_signing) + component(64\_bit)+ component(CFI)$
			\STATE $ expp \leftarrow component(M)/(1+tp)$
            \STATE $ component(metric) \leftarrow expp * component (impact) $
        }
        $ path\_metric \leftarrow path\_metric + component(metric)$ \tcp*{calculate attack surface metric of a path}
    }
    $ TASM \leftarrow TASM + path\_metric$ \tcp*{calculate total attack surface metric}
}
\STATE Return $TASM$
\STATE \}
\end{algorithmic}
\end{algorithm}

\section{Case Study: Electric Power Distribution Control Center}
This section will explore the proposed attack surface metrics on a cyber-physical  electric power distribution system's control center. The system will consist of (i) a simulated distribution power system model, (ii) an AADL model of a control center and SCADA communication, and (iii) an open-source DNP3 software platform (OpenDNP3). The section will also explore different system architectures that can be used to reduce the attack surface of the proposed system by implementing more granular privileges models to improve the isolation of various SCADA system processes. 

\begin{figure*}[!t]
\begin{centering}
\textsf{\includegraphics[width=\textwidth, height = 6 cm]{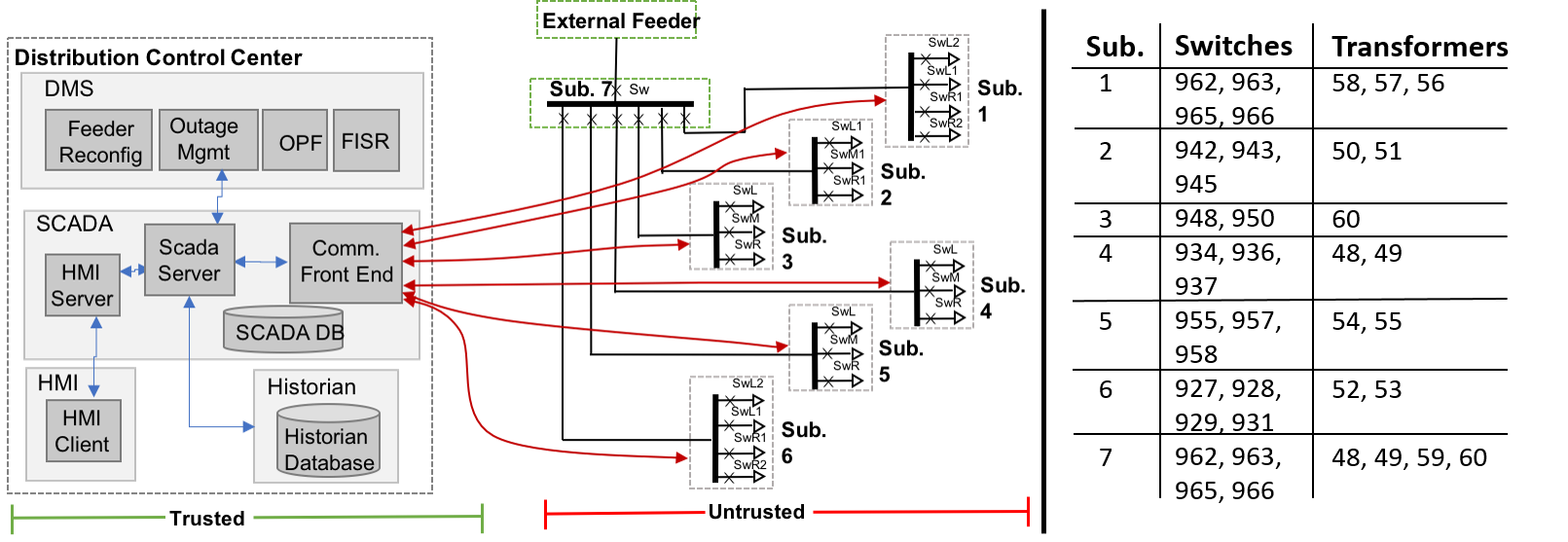}}
\par\end{centering}
\caption{Distribution System Architecture (left) with Controllable Field Devices (right)}
\label{eds-arc}
\end{figure*}

Fig.~\ref{eds-arc} provides an overview of the control center components along with the remote network connectivity to the substations and field devices that monitor and control the grid. Within the control center, there are four subsystems, the SCADA server, human-machine interface (HMI),  distribution management system (DMS), and the historian server/database. The SCADA server performs the communication with a large number of remote devices, including voltage and current transformers, circuit breakers, protection relays, remote switches, transformer taps, and voltage regulators. Typically measurements from these devices are aggregated by a single remote terminal unit (RTU) within a substation and are collected at a 2-4 second interval by the SCADA server and stored in the historian. Furthermore, the SCADA server can also send commands to various devices to control the flow of power through the operation of circuit breakers and the adjustment of transformer tap positions to modify voltage levels. Because the server must maintain a remote communication session with each device, it typically has a front end processor (FEP) process that performs all the SCADA communication with the substation RTUs. The DMS then uses data collected by the SCADA server to execute various algorithms to analysis and optimization grid operation, including power flow, fault identification, and voltage analysis.  The HMI is a workstation used by operators to see system measurements and alarms, while also sending control messages. 

As demonstrated in Fig.~\ref{eds-arc}, this environment can have a broad attack surface due to the high degree of connectivity to remote devices. More specifically, the FEP is the most interconnected process and therefore is the main contributor of the system's attack surface due to its communication with a large number of external systems. If one of these devices is malicious, perhaps due to a previous attack to that device, this connectivity could potentially be used to compromise the SCADA server. Software vulnerabilities that would enable such an attack have already been discovered on popular industry platforms \cite{DNP3vulns}. The remainder of this section will explore both the physical and cyber model used in this case study in more detail. 

\subsection{Physical System Model}\label{phymodel}
\begin{figure*}[!t]
\begin{centering}
\textsf{\includegraphics[width=\textwidth, height= 6 cm]{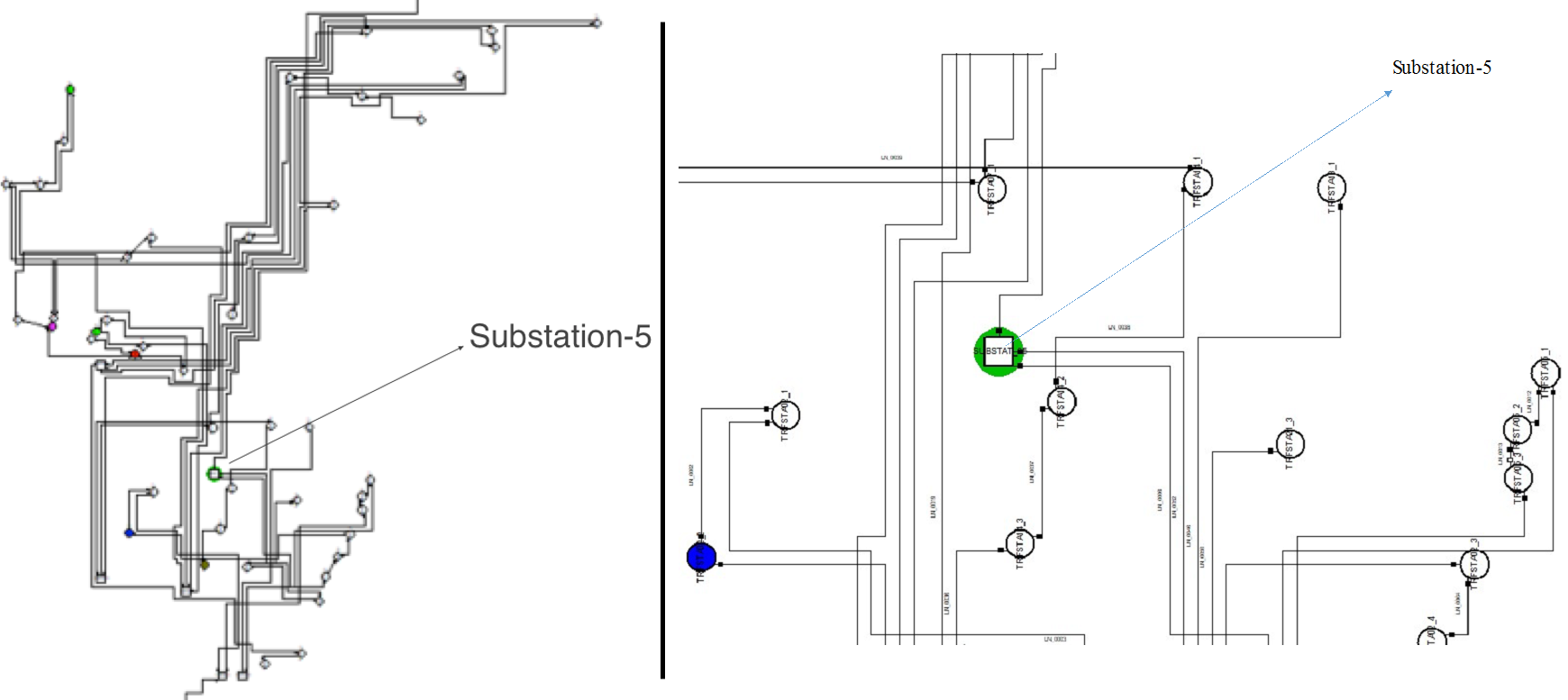}}
\par\end{centering}
\caption{Distribution Feeder Model}
\label{fig:distribution}
\end{figure*}

The physical system for this case-study is a low voltage distribution system model showed in Fig. \ref{fig:distribution} (left). The model includes 7 substations with one external feeder. Fig. \ref{fig:distribution} (right) provides a detailed model of a single substation. There are total 6 feeders at the system, and the total inter-grid power flow is 4744.87 kW. Within the system, there are a number of devices that can be remotely controlled from the control center, including (i) circuit breakers in each substation, (ii) remote switches along certain lines, and (iii) transformer taps changers, each of these are identified in Fig.~\ref{eds-arc}. Circuit breakers are available within each substation, along with multiple breakers or remote switches on the lines connected to other substations or feeders. Therefore, if an attacker can send a malicious disconnect message, they can cause a loss of load to lower feeders. In addition to feeders, the transformer tap changers can also be remotely controlled to help stabilize the voltage.  Transformers in Fig. 3 are connected to the upper feeder which is substation 7 to control the voltage of the system. 

%\begin{table}
%\centering
%\tbl{Controllable devices on substations}{
%\begin{tabular}{| l | l | l |}\hline
%Substation & Substation Switches/Breakers & Transformer Taps \\\hline
%Substation 1 & SW0962, SW0963, SW0965, SW0966 & TRF58, TRF57, TRF56\\
%Substation 2 & SW0942, SW0943, SW0945 & TRF50, TRF51\\
%Substation 3 & SW0948, SW0950 & TRF60\\
%Substation 4 & SW0934, SW0936, SW0937 & TRF48, TRF49\\
%Substation 5 & SW0955, SW0957, SW0958 & TRF54, TRF55\\
%Substation 6 & SW0927, SW0928, SW0929, SW0931 & TRF52, TRF53\\
%Substation 7 & SW0962, SW0963, SW0965, SW0966 & TRF48, TRF49, TRF59, TRF60\\
%\hline
%\end{tabular}}
%\label{controlled}
%\end{table}

%\begin{figure} 
%	\centering
%	\begin{minipage}[b]{0.48\textwidth}
%   	\centering
%    	\includegraphics[width=\textwidth]{figs/full_diagram.png}
%    	\caption{Distribution Feeder Model}
%        \label{fig:full_diagram}
%  	\end{minipage}
%  	\hfill
%  	\begin{minipage}[b]{0.48\textwidth}
%    	\centering
%    	\includegraphics[width=\textwidth]{figs/new_substation5_diagram.pdf}
%    	\caption{System Substation 5 Diagram}
%        \label{fig:sub5_diagram}
%  	\end{minipage}
%\end{figure}

\subsection{Cyber System Model}
\begin{figure}
  \includegraphics[width=0.85\linewidth,height = 4 cm]{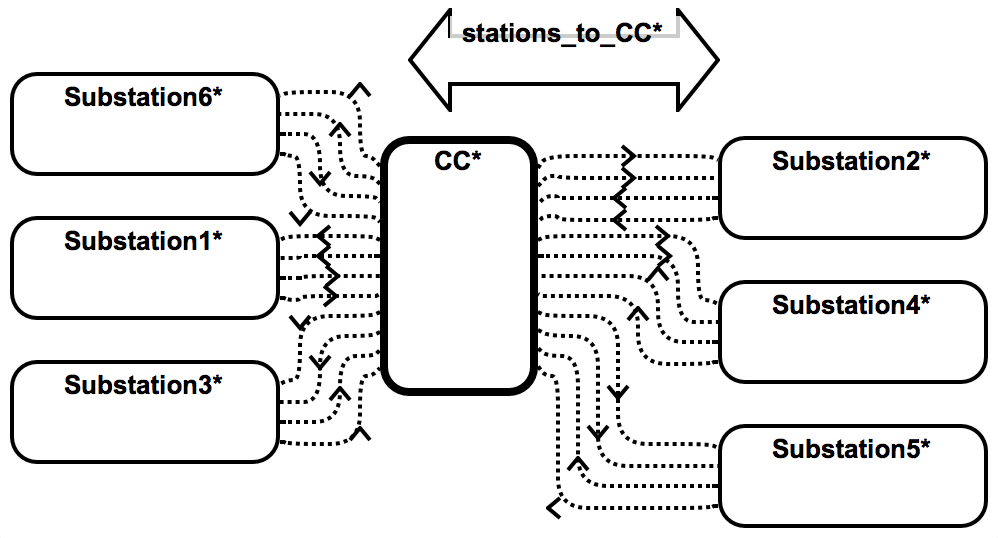}
  \setlength{\abovecaptionskip}{-.0em} 
  \caption{ High-level architecture of the system} 
  \label{fig:highlevel}
\end{figure}

\begin{figure}
  \includegraphics[width=0.8\linewidth,,height = 4 cm]{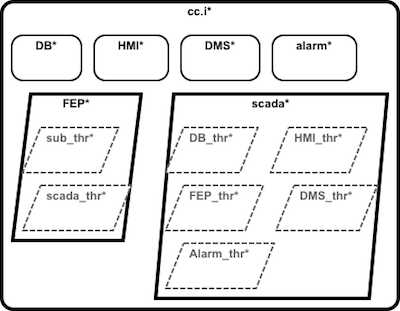}
  \caption{Control Center's components}
  \label{fig:cccomponents}
\end{figure}

\begin{figure}
  \includegraphics[width=0.8\linewidth,height = 7 cm]{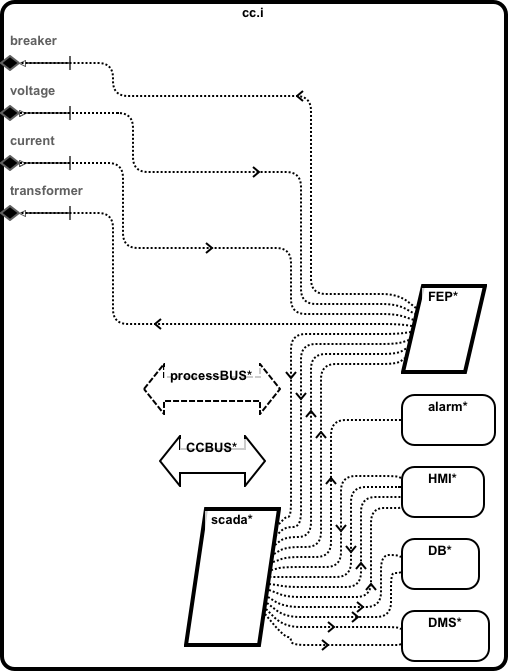}
  \caption{ Control center architecture} 
  \label{fig:cc}
\end{figure}

The control center architecture is modeled using AADL to identify the various processes, buses, subsystems, data, and flows within the cyber system. The high-level architecture including the control center and substation is shown in Fig. \ref{fig:highlevel}. The control center includes a SCADA process, FEP process, DMS, HMI, Database, and alarm processor; all which communicate across a local Ethernet bus. The communication between processes is modeled by a virtual bus. Each process consists of some threads that are responsible for sending or receiving data to/from other components. Fig. \ref{fig:cccomponents} shows each components of control center, while Fig. \ref{fig:cc} shows the broader SCADA communication and control center, where the FEP process communicates across a wide area network bus to all 7 remote substations. Figure \ref{fig:integration-all} in Appendix 1 shows detailed architecture of the system.

%The system also has four kinds of memory defined: Real-Time, Historian, Relay and DMS. The Real-Time memory consists of CT/PT measurement values, the Historian memory consists of tag data values and values from DMS, DMS memory consists of analysis from DMS server and Relay memory consists of Substation data such as CT/PT values. 
Table~\ref{tab:aadldataflow}  defines the data-flows within the architecture that are used to remotely monitor and control all the field equipment in the physical grid model. The first three rows show how telemetry data is read from the field devices in the various substations, into the control center. First, the data transfer from the substations to FEP, then the FEP sends the data to SCADA process. Finally, SCADA process is responsible for sending the data to different parts of control center including HMI, DMS, and database. The last row shows how control messages, which originate at the HMI, are passes through SCADA process and FEP to the substations to actuate the breakers, switches, and transformer taps. 

Furthermore, The communication between substations and FEP are modeled using OpenDNP3, which is a reference implementation of IEEE-1815 (DNP3) protocol and is commonly used to support power grid SCADA communications. The OpenDNP3 libraries are modeled as a subprogram of FEP to communicate with substations. Since OpenDNP3 provides event-oriented communication, it uses two modes (event polling and unsolicited responses) instead of scanning the outstations continuously by the master \cite{opendnpthree}.  In event polling mode, the master asks outstations about changing the data; However, with unsolicited responses, the outstation pushes events to the master when there is a change in the data. Since the number of methods of a process has a direct effect on the exposure of it, the implementation of OpenDNP3 is important. We analyze the OpenDNP3 libraries using the tools such as cflow to find the number of input and output methods are necessary to communicate with the remote devices.

\begin{table*}
\footnotesize
\caption{AADL Data Flows}
\centering
\begin{tabular}{  p{3 cm} | p{5.5 cm} | p{4.5 cm} }\hline	
 Name 	 & 	Flow & Data  				 \\\hline
  Substations[1-6]	to HMI &   $Substations[1-6] \rightarrow FEP \rightarrow SCADA \rightarrow HMI$  & Current [1-6] and Voltage [1-6] Measurement  \\
  Substations[1-6]	to DMS & 	$Substations[1-6] \rightarrow FEP \rightarrow SCADA \rightarrow DMS$ & Current [1-6] and Voltage [1-6] Measurement	 \\
  Substations[1-6] to DB&	$Substations[1-6] \rightarrow FEP \rightarrow SCADA \rightarrow DB$ & Current [1-6] and Voltage [1-6] Measurement	 \\
  HMI to Substations[1-6] &	$ HMI \rightarrow SCADA \rightarrow FEP \rightarrow Substations[1-6]$& Switch Status[1-6], Switch Control[1-6]	 \\
\hline  
\end{tabular}
\label{tab:aadldataflow}
\end{table*}

In addition to the data flows, we also define the system privileges for the SCADA server, which determine what data each process can access. For this model both processes, FEP and SCADA, will possess their own processes defined as $v_{fep}$ and $v_{scada}$ respectively. For each FEP in the system, $v_{fep}$ could be the subset of the data that are presented in Table \ref{results}. For SCADA, $v_{scada}$ is the union of all the $v_{fep}$ of the system.

\subsection{Attack Surface Metrics}
This section will demonstrate the attack surface metrics for the case study system; however, this requires that the trust boundary and attack paths be identified. Furthermore, it requires that the physical impact metrics be computed to assess the risk of an attack to the system. Each of these will be discussed further below.

\subsubsection{Physical Impact}
The physical system impact metrics will be quantified by measuring the loss of load that malicious control messages from control center would have to the grid.  While system telemetry data, such as voltage and current measurements can be manipulated, neither is commonly used for direct feedback control and can only impact the grid by deceiving the operating and encouraging a wrong action. Therefore, this work focuses on the manipulation of the actuation commands to (i) operation the switches and (ii) operated transformer tap positions, as discussed in Section~\ref{phymodel}. The following two sections will discuss the resulting loss of load the control center is compromised, and the attacker can manipulate these messages.

\begin{table}
\footnotesize
\caption{Power Loss After Simulations}{
\label{tab:aadl}
\centering
\begin{tabular}{| l | c | r |}\hline			
  \textbf{Substation}   	&   \textbf{Inner Grid Power}   & 	\textbf{Inner Grid Power}   \\
     	&   \textbf{After Simulation 1}   & 	 \textbf{After Simuation 2}  \\\hline
  Substation 1	& 3164.08 kW, 651 kVar &  3164.08 kW, 651.44 kVar     \\
  Substation 2	& 4431.61 kW, 998.74 kVar & 4441.58 kW, 1040.22 kVar	 \\
  Substation 3  & 4236.29 kW, 927.15 kVar &	4236.29 kW, 927.15 kVar	 \\
  Substation 4	& 4414.58 kW, 1017.84 kVar & 4236.29 kW, 927.15 kVar	 \\
  Substation 5	& 4046.96 kW, 953.37 kVar &	4046.99 kW, 953.46 kVar	 \\
  Substation 6	& 3420.81 kW, 766.86 kVar &	3420.81 kW, 766.86 kVar	 \\
  Substation 7	& 0 kW, 0 kVar & 3906.01 kW, 873.58 kVar	 \\
\hline  
\end{tabular}}
\end{table}

To quantify the loss of load from both malicious switching and transformer taps operations, we simulate the system in steady-state and the operate each switch and tap individually to measure the loss of load from manipulation of that data within either the control center or the SCADA communication to the device.

\begin{table}[t]
\scriptsize
\centering
\caption{Switch Control Data Impact (imp) in KW}{
\begin{tabular}{| lll | lll | lll |}
\hline
Data 	  & Load  & Loss & Data 		& Load  & Loss  & Data & Load  & Loss \\ \hline
 \textbf{Sub1\_All}& \textbf{3164}  & \textbf{1580} & \textbf{Sub3\_All}    & \textbf{4236}  & \textbf{508}   &  Sub5\_SwM   & 4407  & 337  \\ 
Sub1\_SwL1   & 4031  & 713  & Sub3\_SwL  & 4364   & 380  & Sub5\_SwR   & 4716   & 27     \\ \cline{7-9}
 Sub1\_SwR1   & 4123   & 621  &  Sub3\_SwR & 4616    & 128      & \textbf{Sub6\_All}     & \textbf{3420}  & \textbf{1324}  \\ \cline{4-6}
 Sub1\_SwL2    & 4601    & 143    & \textbf{Sub4\_All} & \textbf{4414}  & \textbf{330}  &  Sub6\_SwL1 & 4623            & 121  \\
 Sub1\_SwR2   & 4642  & 102  &  Sub4\_SwL    & 4552  & 192  & Sub6\_SwR1   & 4607  & 137 \\ \cline{1-3}
\textbf{Sub2\_All}    & \textbf{4441}   & \textbf{303}    & Sub4\_SwM    & 4636  & 108    &  Sub6\_SwL2  & 382  & 919      \\
 Sub2\_SwL   & 4600  & 144   &  Sub4\_SwR   & 4715 & 29      &  Sub6\_SwR2  & 4599  & 145                 \\ \cline{4-9}
 Sub2\_SwM   & 4656    & 88    & \textbf{Sub5\_All} & \textbf{4046}  & \textbf{697}       & \textbf{Sub7\_Sw}  & \textbf{0}  & \textbf{4744}  \\
 Sub2\_SwR   & 4674   & 70   &  Sub5\_SwL    & 4412  & 332  &&& \\\hline
\end{tabular}}
\label{results}
\end{table}
    
\textit{Switch Operation:} 
To evaluate the impact of various malicious control center commands, we perform state-state system simulations and then open each switch to calculate the loss of load, emulating the impact of a malicious command to that switch. The results of the switch analysis are displayed in Table~\ref{results}. These results suggest that a malicious operation of switches in Substation 7 will provide the most harmful impact has it controls power flow for the entire feeder and produces a loss of 4744.87 MW. Comparatively, the malicious operation of substation 5's switches have a reduced impact of only 27.91 MW in the case of malicious switch operation. 

\textit{Transformer Tap Operations:} In addition to switch control data, an attacker could also remotely operate transformer tap positions to manipulate the voltage on the feeder. An example is shown for Substation 4 (Fig. \ref{fig:compsubtrans}) targeting transformers $TRF-49$ and $TRF-48$. To negatively impact the voltage, the attacker must have the ability to control both the transformer tap outside of substation 4, along with the one inside substation 4; otherwise, the operator may be able to balance the voltage at the substation. If the voltage deviation of transformers is over plus or minus 5 percent, that transformer is considered having voltage issue. 

If the attacker can manipulate the tap value at substation 4, which is highlighted as red rectangular in Fig. \ref{fig:compsubtrans}, sub transformers outside of substation 4 that are highlighted in the red circle and connected to the substation, have the same over or under voltage issue. Once substation 4 detects over or under voltage issue, it trips the breaker to protect the substation devices. Therefore, $SUBSTAT-04$ at Fig. \ref{fig:compsubtrans} loses all the connection to sub transformers.  To complete the attack, it is assumed that an attacker can control the substation tap changer so that the operator cannot stabilize the voltage.  As a result of this occurrence, the loss of load for transformer tap operation will be equivalent to the breaker operation results displayed in Table \ref{results} except the power loss for substation 7. 

%Fig. \ref{sub7map} shows that substation 7 has 11 power lines, and therefore, has unique results. Only 4 lines have transformers that an attacker can make tap changes, the other 7 lines will still have  power. The loss of power in this case is going to be 508.58 + 330.29 = 838.87 since transformers in substation 4 (\#48, \#49) and 3 (\#59, \#60) are manipulated. 

%\begin{figure}[t]
% \centering
% \includegraphics[width=5in]{figs/sub7map.png}
% \caption{Substation 7 Map}
% \label{sub7map}
%\end{figure}

\begin{figure}
  \includegraphics[width=\linewidth]{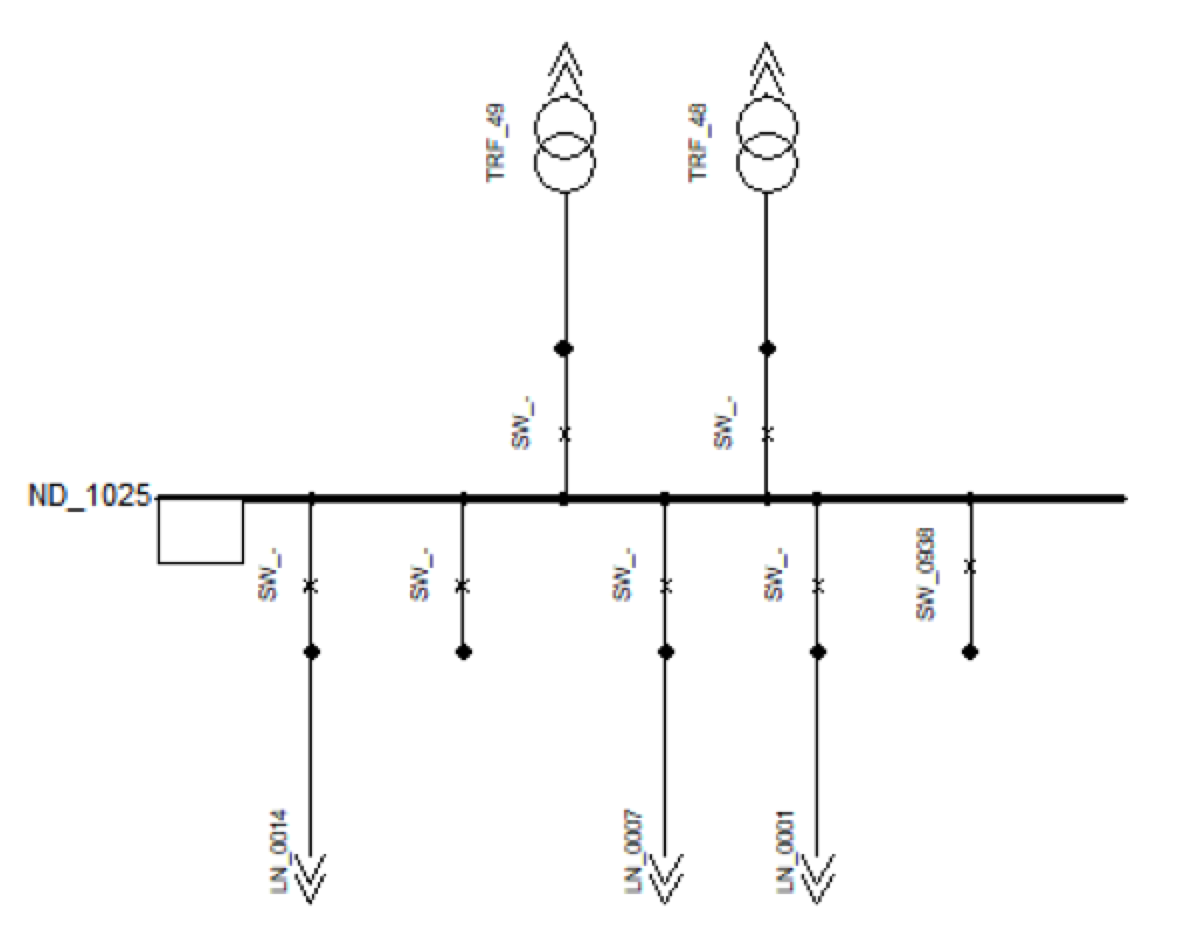}
 		\caption{Substation 4 Map}
 		\label{fig:asubmap}
\end{figure}
\begin{figure}
  \includegraphics[width=\linewidth, height = 7 cm]{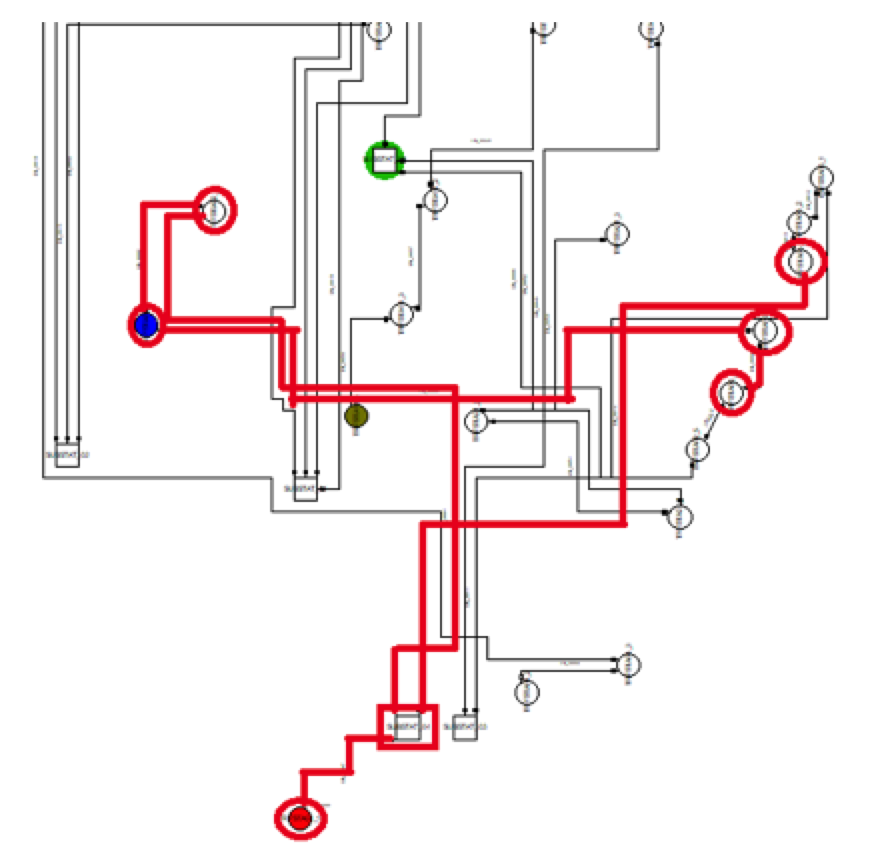}
 		\caption{Compromised Substation 4 and Sub Transformers}
 		\label{fig:compsubtrans}
\end{figure}

\subsubsection{Cyber Metrics}
This section will demonstrate the attack surface analysis for the cyber model, including the OpenDNP3 software subprogram. The $AP_s$ for the system includes 6 attack paths from substations to FEP process. To demonstrate how different attack surfaces metrics will be produced based on different system features and properties (as shown in Table~\ref{tab:conexp}) explores how different values for each variable produces a unique attack surface score ($TASM$). Therefore, different values of the variables for $ASLR$, $DEP$, $code\_sign$, $64\_bit$, and $CFI$ have been assigned to explore their attack surface impact. For each of these mechanisms, if it is applied the value is 1 (True). Otherwise, zero (false) is assigned to the mechanism. We investigate different case studies with different values for the mechanisms.

\textit{OpenDNP3 Subprogram:} The OpenDNP3 program is analyzed to identify the number of input/output methods ($M$). To analyze the number of methods in OpenDNP3, we use the GNU cflow program, which analyzes a collection of C source files and generates outputs in control flow graph format \cite{cflow}. By analyzing DNP3 libraries, we identify that there are 14 methods applied for sending and receiving data, so $M=14$. Furthermore, this program will be used to communicate with remote substations, so we define various values for the $encryption\_algorithm$, $key\_exchange$, $hash\_function$, and $media\_type$.

\subsubsection{Total Attack Surface Metrics} 
In this part, we investigate the calculation of the $TASM$ metric on this proposed test system while varying a number of security properties. In the first case, we assume that the control center only has one FEP whose privileges, $v_{fep}=D$, has complete access to all control data for the substation switches. For this privilege and data model, we define 5 different sub-scenarios varying the security mechanism of substations and FEP. In the first scenario, we evaluate attack surface metric with the highest level of connection and process security. Then, in the second scenario, we apply the highest connection security, but without process security enabled. In the third scenario, we calculate the metric with the highest process security mechanism without any connection security implemented. In the fourth scenario, we evaluate the system without any security mechanism for both the connection and process. Finally, In the fifth scenario, we apply a random security mechanism. The results show that the total attack surface metric is the lowest level when we use the highest security mechanism for the connection and FEP. Moreover, we can find that the process security mechanism is more significant than the connection security mechanism by comparing the second and third scenarios.

In other case studies, the attack surface metric is evaluated for the system where the FEP is divided into two processes with separate privileges ($v_{fep1}, v_{fep2}$), such that an attack to one process has a limited ability to manipulate data to only the data of the devices in that row. We define security properties for each process similar to the approach used in the first case study. We allocate high-security properties to one process and assign the other one low-security properties. Each device communicates with one of the processes depend on the data that they transfer. If the process needs improved security, they interact with the FEP with high-security properties. Otherwise, They communicate with the FEP with low-security properties. Table \ref{tab:totalvaluemultiFEP} shows different case studies. In each case study, the $FEP1$ has more security mechanism compared with $FEP2$ except two last case studies which have same security mechanism.

In the second case study, the first four substations that have the highest security mechanism communicate with $FEP1$ and the substations 5 and 6 communicate with $FEP 2$. In this case study, $FEP1$ has the highest security mechanism and $FEP 2$ has no security mechanism. In the third and fourth case studies, substations 1 and 6 communicate with $FEP 1$ and substations 2 to 5 communicate with $FEP 2$. There are two differences between these two case studies. In the third case study, substations 2 to 5 do not have any security mechanism. However, In the fourth case study, they have highest security mechanism. Moreover, In the third case study, $FEP 2$ does not have any security mechanism. However, in the fourth case study, it supports code signing and CFI. The case studies 5 and 6 show the condition in which some switches of each substation communicate with $FEP1$ and other switches of substation communicate with $FEP2$. For example, switch $L1$ and switch $L2$ of substation 1 communicate with $FEP1$ and switch $R1$ and switch $R2$ of substation 1 communicate with $FEP2$ in the sixth case study. In the two last case studies, we show the cases that both FEPs have the highest security mechanism and substations have the highest connection security mechanism. The results show that the $TASM$ is smaller in these cases compared with the first scenario of the first case study where we have only one FEP.   

\begin{table*}
\caption{Total Attack Surface value for different case studies}{
\label{tab:totalvaluemultiFEP}
\scriptsize
\centering
\begin{tabular}{|p{.5cm}|p{1cm} p{.5cm} p{.5cm} c c c|p{.5cm} p{.7cm} p{.75cm} p{.83cm}|p{.25cm} p{.2cm} p{.55cm} p{.3cm} p{.15cm}|r|}\hline
\textbf{\makecell{Priv. \\($V(s)$)}}&\multicolumn{6}{|c|} {\textbf{Data ($D$)}} & \multicolumn{4}{|c|}{\textbf{Connection  Parameters ($exp_c$)}}  &\multicolumn{5}{|c|} {\textbf{Process  Parameters ($exp_p$)}} &\textbf{\makecell{Total\\Attack\\surface\\metric\\($TASM$)}} \\ \hline
&Sub1&Sub2&Sub3&Sub4&Sub5&Sub6&enc.&key\_ex.&hash\_func.&media\_type&ASLR&DEP&code\_sig.&64\_bit&CFI& \\ \hline
         
&&&&&&&AES&2048bits&SHA-2&wired&Y&Y&Y&Y&Y&\num[group-separator={,}]{13609} \\
&&&&&&&NULL&NULL&NULL&wireless&Y&Y&Y&Y&Y&\num[group-separator={,}]{39500} \\ 
FEP1&L1,L2,R1,R2&L,M,R&L,R&L,M,R&L,M,R&L1,L2,R1,R2&AES&2048bits&SHA-2&wired&N&N&N&N&N&\num[group-separator={,}]{68948} \\ 
&&&&&&&NULL&NULL&NULL&wireless&N&N&N&N&N&\num[group-separator={,}]{94840} \\ 
&&&&&&&3DES&1024bits&SHA-1&wireless&Y&N&Y&N&Y&\num[group-separator={,}]{20580} \\ \hline
FEP1&L1,L2,R1,R2&L,M,R&L,R&L,M,R&&&AES&2048bits&SHA-2&wired&Y&Y&Y&Y&Y&\num[group-separator={,}]{39655} \\ 
FEP2&&&&&L,M,R&L1,L2,R1,R2&NULL&NULL&NULL&wireless&N&N&N&N&N& \\ \hline
FEP1&L1,L2,R1,R2&&&&&L1,L2,R1,R2&AES&2048bits&SHA-2&wired&Y&Y&Y&Y&Y&\num[group-separator={,}]{40373} \\ 
FEP2&&L,M,R&L,R&L,M,R&L,M,R&&NULL&NULL&NULL&wireless&N&N&N&N&N& \\ \hline
FEP1&L1,L2,R1,R2&&&&&L1,L2,R1,R2&AES&2048bits&SHA-2&wired&Y&Y&Y&Y&Y&\num[group-separator={,}]{16534} \\ 
FEP2&&L,M,R&L,R&L,M,R&L,M,R&&AES&2048bits&SHA-2&wired&N&N&Y&N&Y& \\ \hline
FEP1&R1,L1&L&L&L,M&L,M&L2&AES&2048bits&SHA-2&wired&Y&Y&Y&Y&Y&\num[group-separator={,}]{15908} \\       
FEP2&R2,L2&M,R&R&R&R&L1,R1,R2&AES&2048bits&SHA-2&wired&N&N&Y&N&Y& \\ \hline
FEP1&L1,L2&L,M&L&L&L,M&L2,R2&AES&2048bits&SHA-2&wired&Y&Y&Y&Y&Y&\num[group-separator={,}]{16734} \\       
FEP2&R1,R2&R&R&M,R&R&L1,R1&AES&2048bits&SHA-2&wired&N&N&Y&N&Y& \\ \hline
FEP1&L1,L2,R1,R2&&&&&L1,L2,R1,R2&AES&2048bits&SHA-2&wired&Y&Y&Y&Y&Y&\num[group-separator={,}]{24829} \\ 
FEP2&&L,M,R&L,R&L,M,R&L,M,R&&AES&2048bits&SHA-2&wired&Y&Y&Y&Y&Y& \\ \hline
FEP1&L1,L2&L,M&L&L&L,M&L2,R2&AES&2048bits&SHA-2&wired&Y&Y&Y&Y&Y&\num[group-separator={,}]{13592} \\       
FEP2&R1,R2&R&R&M,R&R&L1,R1&AES&2048bits&SHA-2&wired&Y&Y&Y&Y&Y& \\ \hline
\end{tabular}}
\end{table*}

\section{Conclusions}
Modern cyber-physical systems are growing increasingly complex and interconnected, which expands their attack surface to remote cyber intrusions. While a number of research efforts have explored techniques to protect these systems from false data attacks, there has been limited work research in understanding the risk to remote exploitations and intrusions resulting from this connectivity. This paper introduces attack surface analysis metrics and algorithms for cyber-physical systems that incorporate physical system impact metrics, along with a variety of cyber factors across system security, communications security, and software complexity. Software has been developed to automate this analysis through AADL-based models to enable the analysis to perform on an arrange real industry CPS projects. Furthermore, a case study has been demonstrated on a cyber-physical distribution grid system models, which include a cyber model consisting of an AADL model for a control center and an OpenDNP3 software library, along with a physical distribution power grid with 7 substations. 
%\end{document}  % This is where a 'short' article might terminate

\appendix
%Appendix A
\section{AADL Model}
\begin{figure*}[h!]
\centering
\includegraphics[max height=23 cm,max width=18cm]
{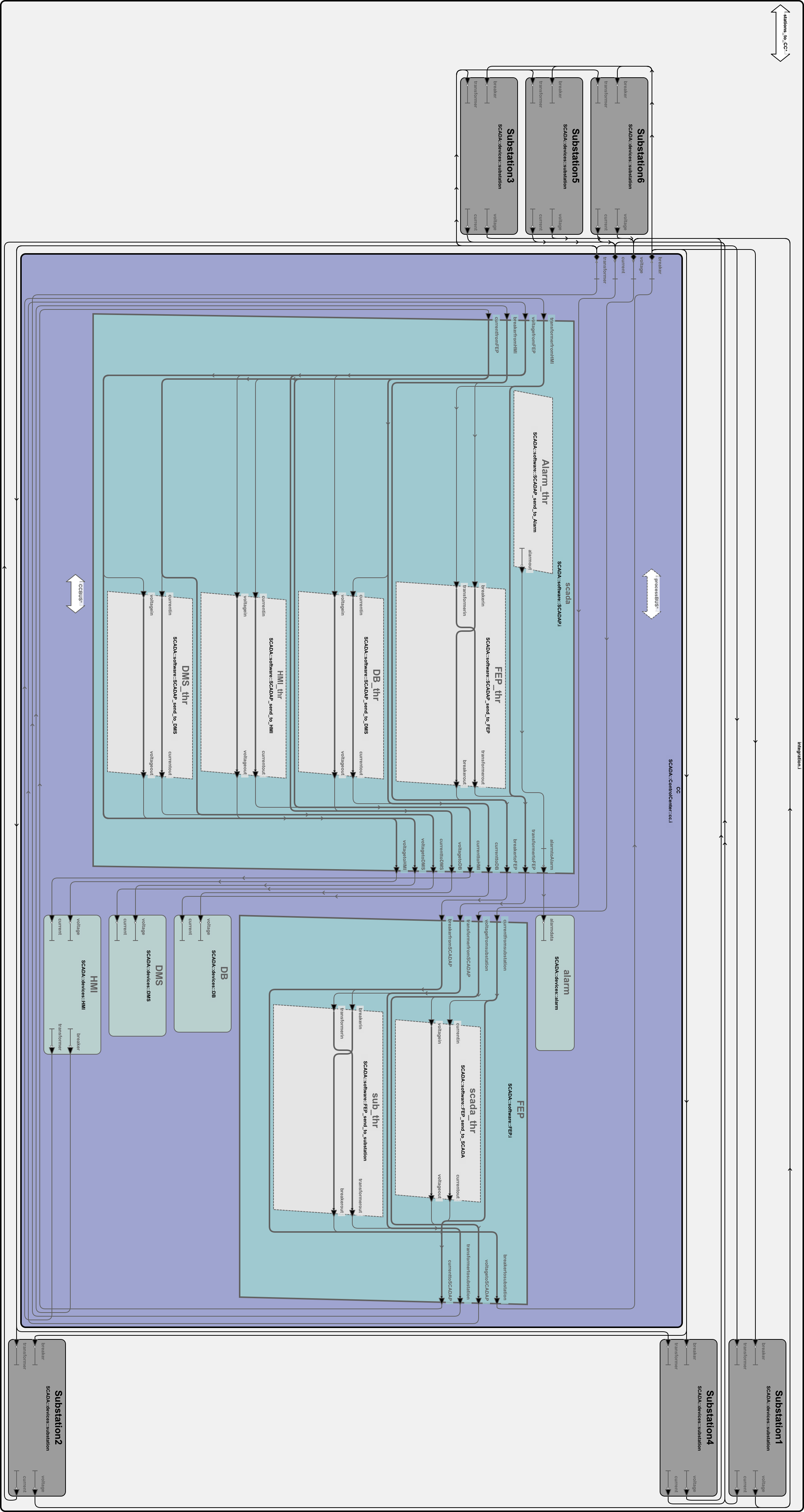}
\caption{Detail Architecture of the system}
\label{fig:integration-all}
\end{figure*}

In this part, we explain the AADL model that used for analyzing the attack surface. Seven substations communicate to control center through a bus. Each substation has four ports that transfer voltage, current, breaker, and transformer between the control center and them. Control center has same ports for transferring the data. The first part of the control center is a front-end process(FEP) that is responsible for communicating to the substations. FEP send and receive data to/from the SCADA process. SCADA process is the main process in the control center. It receives the data from FEP and sends them to Database, DMS, and HMI. Moreover, It processes the data and sends an alarm signal if it is needed. SCADA process is also responsible for sending the command from HMI to FEP. SCADA communicates with other parts of control center using an internal bus. Figure \ref{fig:integration-all} shows the AADL detailed model.
% \begin{acks}
%   The authors would like to thank Dr. Yuhua Li for providing the
%   MATLAB code of the \textit{BEPS} method.

%   The authors would also like to thank the anonymous referees for
%   their valuable comments and helpful suggestions. The work is
%   supported by the \grantsponsor{GS501100001809}{National Natural
%     Science Foundation of
%     China}{http://dx.doi.org/10.13039/501100001809} under Grant
%   No.:~\grantnum{GS501100001809}{61273304}
%   and~\grantnum[http://www.nnsf.cn/youngscientists]{GS501100001809}{Young
%     Scientists' Support Program}.

% \end{acks}